\documentclass[11,runningheads,a4paper]{llncs}

\usepackage[sort&compress,numbers]{natbib}
\usepackage{comment}

\usepackage{amssymb,amsmath}
\usepackage{graphicx}

\usepackage{url}
\urldef{\mailsa}\path|beka@cmu.edu, {sventura, msadinle, fienberg}@stat.cmu.edu|

\usepackage{arabtex}
\usepackage{utf8}
\usepackage{soul}

\usepackage[switch]{lineno}
\RequirePackage[colorlinks,citecolor=blue,urlcolor=blue]{hyperref}
\usepackage[ruled,lined]{algorithm2e}
\SetKw{KwSet}{Set}

\begin{document}

\mainmatter  

\title{Probabilistic Blocking with An Application to the Syrian Conflict}

\titlerunning{Probabilistic Blocking with An Application to the Syrian Conflict}

\author{Rebecca C. Steorts$^{1}$ and Anshumali Shrivastava$^{2}$%
 }

\authorrunning{Steorts and Shrivastava}

\institute{$^{1}$Department of Statistical Science, affiliated faculty, Computer Science, Biostatistics and Bioinformatics, the information initiative at Duke (iiD), and the Social Science Research Institute (SSRI) Duke University \\
$^{2}$Department of Computer Science, Rice University 
\path|{beka}@stat.duke.edu; {anshumali}@rice.edu, |
}
\maketitle

\begin{abstract}
Entity resolution seeks to merge databases as to remove duplicate entries where unique identifiers are typically unknown. We review modern blocking approaches for entity resolution, focusing on those based upon locality sensitive hashing (LSH). First, we introduce $k$-means locality sensitive hashing (KLSH), which is based upon the information retrieval literature and  clusters similar records into blocks using a vector-space representation and projections. Second, we introduce a subquadratic variant of LSH to the literature, known as Densified One Permutation Hashing (DOPH). Third, we propose a weighted variant of DOPH.  We illustrate each method on an application to a subset of the ongoing Syrian conflict, giving a discussion of each method. 
\end{abstract}

\section{Introduction}
\label{sec:intro}
A commonly encountered problem in statistics, computer science, and machine learning is merging noisy data sets that contain duplicated entities, which is known as entity resolution (record linkage or de-duplication). Entity resolution tasks are intrinsically difficult because they are quadratic in computational complexity. In addition, for such tasks to be accurate, one often seeks models that are robust to model-misspecification and also have low error rates (precision and recall). These criteria are both difficult to satisfy, and have been at the core of entity resolution research \cite{christen_2012, Herzog_2007, Herzog:2010, winkler06overview}. 

One way of approaching the computational complexity barrier is by partitioning  records into ``blocks'' and treating records in different blocks as non-co-referent {\em a priori} \cite{christen_2012, Herzog_2007}. There are several techniques for constructing a blocking partition. The most basic method picks certain fields (e.g., geography, or gender and year of birth) and places records in the same block if and only if they agree on all such fields.  This amounts to an {\em a priori} judgment that these fields are error-free. This is known as traditional blocking, and is a deterministic scheme. 
Unlike traditional blocking, probabilistic schemes such as locality sensitive hashing (LSH) use all the fields of a record, and can be adjusted to ensure that blocks are manageably small.  For example, \cite{christen_2014} introduced data structures for sorting and fast approximate nearest-neighbor look-up within blocks produced by LSH.  

This approach is fast and has high recall (true positive rate), but suffers from low precision (too many false positives).  In addition, this approach is called \emph{private} if, after the blocking is performed, all candidate records pairs are  compared and classified into matches/non-matches using computationally intensive ``private" comparison and classification techniques, e.g., see \cite{christen_2009}. 

LSH has been recently proposed as a way of blocking for entity resolution, where one place similar records into bins or blocks. LSH methods are defined by a type of similarity and a type of dimension reduction \citep{broder_1997}. 
Recently, \cite{steorts14comparison} proposed clustering-based blocking schemes ---  $k$-means locality sensitive hashing (KLSH), which is based upon the information retrieval literature and  clusters similar records into blocks using a vector-space representation and projections. (While KLSH had been used before within the information retrieval literature, this is the first known instance of its application to entity resolution \citep{pauleve_2010}). \cite{steorts14comparison} showed that KLSH  gave improvements over popular methods in the literature such as traditional blocking, canopies \citep{mccallum_2000}, and k-nearest neighbors clustering.  In addition, \cite{shrivastava2014densifying} showed that minwise hashing based approaches are superior to random projection based approaches when the data is very sparse and feature poor. Furthermore, computational improvements can be gained by using the recently proposed densification scheme known as densified one permutation hashing (DOPH)~\citep{shrivastava2014densifying,shrivastava2014improved}.  Specifically, the authors proposed an efficient substitute for minwise hashing, which only requires one permutation (or one hash function) for generating many different hash values needed for indexing. In short, the algorithm is linear (or constant) in the tuning parameters, making this algorithm very computationally efficient. 

In this paper, we review traditional blocking methods that are deterministic, and describe why such methods are not practical. We then review scalable LSH methods for blocking.  Specifically, we give two recent approaches an methods from above that are scalable to large entity resolution data sets -- KLSH and DOPH. Since both methods are known to work well on toy examples, we illustrate both algorithms on a real data set taken from a subset of the Syrian conflict, which is likely to be more realistic of industrial sized data sets. 
We illustrate evaluation metrics of all methods and the computational run time. 


\section{Blocking Methods}
\label{sec:block}
Blocking is a computational tool used in entity resolution that allows one to place similar records into blocks or partitions using either a deterministic or probabilistic mechanism. We first review traditional blocking methods, and then review probabilistic blocking methods. We propose two probabilistic blocking methods for large scale blocking methods based upon LSH.
%

\subsection{Traditional Blocking}
\label{sec:traditional}

Traditional blocking requires domain knowledge to
pick out highly reliable, if not error-free, fields for blocking.  While traditional blocking is intuitive and easy to implement, it has at least four drawbacks.  The first one is that the resulting blocks may still be so large that linkage within them is computationally impractical.  The second is that because blocks {\em only} consider selected fields, much time may be wasted comparing records which happen to agree on those fields but are otherwise radically different. The third is due to the fact that traditional blocking methods are by nature deterministic, and thus, must be changed for each application at hand. The fourth is that a deterministic method cannot be guaranteed to be private. Given that traditional blocking is impractical for many reasons, we refer readers to \cite{steorts14comparison}, and we focus instead on probabilistic types of blocking, namely variants LSH. 

\subsection{Variants of Locality Sensitive Hashing}
\label{sec:lsh}
In this section, we first provide terminology, known as shingling, that is essential for using LSH for blocking. Next, we describe how can one produce blocks using k-means LSH (KLSH). Then we introduce the notation of LSH, and the linear variant --- Densified One Permutation Hashing (DOPH). Finally, we propose a weighted variant of DOPH (weighted DOPH). 

\subsection{Shingling}
In entity resolution tasks, each record can be represented as a string of textual information. It is often convenient to represent each record instead by a ``bag", ``shingle" 
(or ``multi-set'') of length-$k$ contiguous sub-strings that it contains. In this paper, we use a k-shingle (k-gram or k-token) based approach to transform the records, and our representation of each record is a set, which consists of all the $k$-contiguous characters occurring in record string. 

As an illustration, for the record \text{BAKER, TED}, we separate it into a 2-gram representation. The resulting set is the following: $$\text{BA, AK, KE, ER, ER, TE, ED}.$$
For example, consider \text{Sammy, Smith}, whose 2-gram set representation is $$\text{SA, AM, MM, MY, MS, SM, MI, IT, TH}.$$ We now have two records that have been transformed into a 2-gram representation. Thus, for every record (string) we obtain a set $\subset \mathcal{U}$, where the universe $\mathcal{U}$ is the set of all possible $k$-contiguous characters.

\subsection{KLSH}
We explore a simple random projection method, KLSH, where the similarity between records is measured using the inner product of a bag-of-shingled vectors that are weighted by their inverse document frequency. We first construct a k-shingle of a record by replacing the record by a bag (or multi-set) of length-k contiguous sub-strings that it contains. After the shingles are created, the 
 dimensionality of the bag-of-shingled vectors is then reduced using random projections and by clustering the low-dimensional projected vectors via the k-means algorithm. That is, the mean number of records per cluster is controlled by $n/c,$ where $n$ is the total number of records and $c$ is the number of block-clusters.

\subsection{Locality Sensitive Hashing}
We now turn to LSH, which is used in computer science and database engineering as a way of rapidly finding approximate nearest neighbors
\citep{Proc:Indyk_STOC98,gionis_1999}. Specifically, the variant of LSH that we utilize is scalable to large databases, and allows for similarity based sampling of entities in a subquadratic amount of time.

In LSH, a hash function is defined as $y = h(x),$ where $y$ is the \emph{hash code} and $h(\cdot)$ the \emph{hash function}. A \emph{hash table} is a data structure that is composed of \emph{buckets} (not to be confused with blocks), each of which is indexed by a \emph{hash code}. Each reference item (record) $x$ is placed into a bucket $h(x).$

More precisely, LSH is a family of function that map vectors to a discrete set, namely, $h:\mathbb{R}^D \rightarrow \{1, \ 2,\cdots , M\}$, where $M$ is in finite range. Given this family of functions, similar points (records) are likely to have the same hash value compared to dissimilar points (records). The notion of similarity is specified by comparing two vectors of points (records), $x$ and $y.$ We will denote a general notion of similarity by $\text{SIM}({x,y}).$ In this paper, we only require a relaxed version LSH, and we define this below. For a complete review of LSH, we refer to \cite{rajaraman_2012}. Formally, a LSH is defined by the following definition below:

\begin{definition} (Locality Sensitive Hashing (LSH))\ Let $x_1, \ x_2, \ y_1, \ y_2 \in \mathbb{R}^D$ and suppose $h$ is chosen uniformly from a family $\mathcal{H}.$ Given a similarity metric, $\text{SIM}(x,y)$, $\mathcal{H}$ is locality sensitive if $\text{SIM}(x_1,x_2)\ge Sim(y_2,y_3)$ then ${Pr}_\mathcal{H}(h(x_1) = h(x_2)) \ge {Pr}_\mathcal{H}(h(y_1) = h(y_2)),$ where ${Pr}_\mathcal{H}$ is the probability over the uniform sampling of $h$. 
\end{definition}



\subsubsection{Minhashing}
One of the most popular forms of LSH is minhashing~\citep{Proc:Broder}, which has two key properties --- a type of similarity and a type of dimension reduction. The type of similarity used is the Jaccard similarity and the type of dimension reduction is known as the minwise hash, which we now define.

Let $\{0,1\}^D$ denote the set of all binary $D$ dimensional vectors, while $\mathbb{R}^D$ refers to the set of all $D$ dimensional vectors (of records). Note that records can be represented as a binary vector (or set) representation via a shingling representation
%
More specifically, given two record sets (or equivalently binary vectors) $x,y \in \{0,1\}^D,$ the Jaccard similarity between $x, y \in \{0,1\}^D$  is
$
\mathcal{J} = \dfrac{|x \cap y|}{|x \cup y|},
$
where $|\cdot|$ is the cardinality of the set.

More specifically, the minwise hashing family applies a random permutation $\pi$, on the given set $S$, and stores only the minimum value after the permutation mapping, known as the \emph{minhash}.  Formally, the minhash is defined as $h_{\pi}^{min}(S) = \min(\pi(S))$, where $h(\cdot)$ is a hash function.

Given two sets $S_1$ and $S_2$, it can be easily shown that
\begin{equation}
\label{eq:MinHash}
Pr_{\pi}({h_{\pi}^{min}(S_1) = h_{\pi}^{min}(S_2)) =  \frac{|S_1 \cap S_2|}{| S_1 \cup S_2|}},
\end{equation}
where the probability is over uniform sampling of $\pi$. It follows from Equation~\ref{eq:MinHash} that minhashing is a LSH family for the Jaccard similarity.\footnote{In this paper, we utilize a shingling based approach, and thus, our representation of each record is likely to be very sparse. Moreover, \cite{shrivastava2014defense} showed that minhashing based approaches are superior compared to random projection based approaches for very sparse data sets.}

\subsection{DOPH}
In this section, we introduce the linear variant of LSH, known as DOPH. Let $K$ be the number of hash functions and let $L$ be the number of hash tables. 
A  $(K,L)$ parameterized blocking scheme requires $K \times L$ hash computations per record. For a single record, this requires storing and processing hundreds (or even thousands) of very large permutations.  This in turn requires hundreds or thousands of passes over each record. Thus, traditional minwise hashing is prohibitively expensive for large or moderately sized data sets. In order to cross-validate the optimal $(K,L)$ tuning parameters, we need multiple independent runs of the $(K,L)$ parameterized blocking scheme. This expensive computation is a major computational concern.
To avoid this computational issue, we can use \emph{one permutation} of the hash function, where $k = K \times L$ minhashes are made in one single pass over the data \cite{shrivastava2014densifying,shrivastava2014improved}. 

Due to sparsity of data vectors (from shingling), empty blocks (in the hash tables) are possible and destroy LSH's essential property \citep{rajaraman_2012}.
To restore this, 
we rotate the values of non-empty buckets and assign a number to each of the empty buckets. 
Our $KL$ hashed values are simply the final assigned values in each of the $KL$ buckets. The final values were shown to satisfy Equation~\ref{eq:MinHash}, for any $S_1, \ S_2$, as shown in~\citep{shrivastava2014densifying,shrivastava2014improved}.

\subsection{Weighted DOPH}
\label{sec:doph}

Minhashing, however, only uses the binary information and ignores the weights (or values) of the components, which is important for entity resolution problems due to the unbalanced nature of the data (small amount of duplicate records). This is the reason why we observe slightly better performance for synthetic data of LSH methods used in \cite{steorts14comparison}, one of which is based upon random projections. To explore this more broadly, we examine the power of minwise hashing for entity resolution, a situation where the data is quite unbalanced, while simultaneously utilizing the weighting of various components.

Suppose now $x, y$ are non-negative vectors. For our problem, we are only interested in non-negative vectors because shingle based representations are always non-negative. We utilize the generalization of Jaccard similarity for real valued vectors in $\mathbb{R}^D$, Unlike the minhash, this variant is sensitive to the weights of the components, and is defined as
\begin{align}
 \mathcal{J}_w = \frac{\sum_i \min\{x_i,y_i\}}{ \sum_i \max\{x_i,y_i\}}
 = 1 - \frac{\| x - y \|_1} { \sum_i \max\{x_i,y_i\}},
\end{align}
where $||\cdot||_1$ represents the $\ell_1$ norm.
Consistent weighted sampling~\citep{charikar2002similarity, gollapudi2006exploiting, manasse2010consistent, ioffe2010improved} is used for hashing the weighted Jaccard similarity $J_w$. In our application to the subset of the Syrian dataset,  we find minhash and weighted minhash give similar error rates, which can be seen in \S \ref{sec:app}.


With DOPH,  the traditional minwise hashing scheme is linear or constant in the tuning parameters.  
For the weighted version of minhashing, we propose a different way of generating hash values for weighted Jaccard similarity, similar to that of~\cite{charikar2002similarity, gollapudi2006exploiting}. 
As a result, we obtain the fast and practical one pass hashing scheme for generating many different hash values with weights, analogous to DOPH for the unweighted case. Overall, we require only one scan of the record and only one permutation.

Given any two vectors $x,y \in \mathbb{R}^D$ as the shingling representation,
we seek hash functions $h(\cdot)$, such that the collision probability between two hash functions is small. More specifically, this means that
\begin{equation}Pr(h(x) = h(y)) = \frac{\sum_i {\min\{x_i,y_i\}}}{\sum_i\max\{x_i,y_i\}}.
\end{equation}

Let $\delta$ be a quantity such that all components of any vector $x_i = I_i^x\delta$ for some integer $I_i^x$.\footnote{The assumption holds when dealing with floating point numbers for small enough $\delta$.}  Let  the maximum possible component $x_i$ for any record be $x$  and let $M$ be an integer such that $x_i = M \delta$. Thus, $\delta$ and $M$ always exist for finitely bounded data sets over floating points.

Consider the transformation  $T: \mathbb{R}^D \rightarrow \{0,1\}^{M\times D}$, where for $T(x)$ we expand each component $x_i = I\delta$ to $M$ dimensions and with the first $I$ dimensions have value $1$ and the rest value $0$.

Observe that for vectors $x$ and $y$, $T(x)$ and $T(y)$ are binary vectors and
\begin{align}
&\frac{|T(x) \cap T(y)|}{|T(x) \cup T(y)|} = \frac{\sum_i \min\{I_i^x,I_i^y\}}{ \sum_i \max\{I_i^x,I_i^y\}} \label{eqn:anshu} \\
&= \frac{\sum_i \min\{I_i^x,I_i^y\} \delta}{ \sum_i \max\{I_i^x,I_i^y\} \delta} = \frac{\sum_i {\min\{x_i,y_i\}}}{\sum_i\max\{x_i,y_i\}} \notag
\end{align}
In other words, the usual resemblance (or Jaccard similarity) between the transformed $T(x)$ and $T(y)$ is precisely the weighted Jaccard similarity between $x$ and $y$ that we are interested in.
Thus, we can simply use the DOPH method of~\cite{shrivastava2014densifying,shrivastava2014improved} on $T(x)$ to get an efficient LSH scheme for weighted Jaccard similarity defined by Equation~\ref{eqn:anshu}.  The complexity here is $O(KL+ \sum_i I_i  )$ for generating $k$ hash values, a factor improvement over $O(k\sum_i I_i)$ without the densified scheme.

Often $I_i$ is quite large (when shingling) and $\sum_i I_i$ is large as well.
When $\sum_i I_i$  is large, \cite{gollapudi2006exploiting} give simple and accurate  approximate hashes for weighted Jaccard similarity.  They divide all components $x_i$ by a reasonably big constant so that $x_i \le 1$ for all records $x$. After this normalization, since $x_i \ge 0$, for every $x,$ we generate another bag of word $x_S$ by sampling each $x_i$ with probability $x_i \le 1$. Then $x_S$ is a set (or binary vector) and for any two records $x$ and $y$, the resemblance between $x_S$ and $y_S$ sampled in this manner is a very accurate approximation of the weighted Jaccard similarity between $x$ and $y$. After applying the DOPH scheme to the shingled records, we generate $k$ different hash values of each record in time $O(KL+d)$, where $d$ is the number of shingles contained in each record. This is a vast improvement over $O(KL+ \sum_i I_i  )$.  Algorithm~\ref{alg:hashgeneration} summarizes our method for generating $k$ different minhashes needed for blocking.

\begin{algorithm}[h]
\caption{Fast $KL$ hashes}
\label{alg:hashgeneration}
\KwData{record $x$, }
\KwResult{$KL$ hash values for blocking}
 $x_S = \phi$\;
\ForAll{$x_i > 0$}{
 $x_S \cup i$ with probability proportional to $x_i$\;
}
\Return $KL$ densified one permutation hashes (DOPH) of $x_S$
\end{algorithm}

\section{Evaluation Metrics}
We evaluate each of our hashing methods below using recall and reduction ratio (RR). The recall measures how many of the actual true matching record pairs have been correctly classified as matches. There are four possible classifications. 
First, record pairs can be linked under both the truth and under the estimate, which we refer to as \emph{correct links} (CL). Second, record pairs can be linked under the truth but \emph{not} linked under the estimate, which are called
\emph{false negatives} (FN).
Third, record pairs can be \emph{not} linked under the truth but linked under the estimate, which are called
\emph{false positives} (FP).
Fourth and finally, record pairs can be \emph{not} linked under the truth and also \emph{not} linked under the estimate, which we refer to as \emph{correct non-links} (CNL). 
The vast majority of record pairs are classified as correct non-links in most practical settings.
Then the true number links is $\text{CL}+\text{FN}$, while the estimated number of links is $\text{CL}+\text{FP}$.  The usual definitions of false negative rate and false positive rate are
\begin{equation}\notag
\text{FNR}=\frac{\text{FN}}{\text{CL+FN}},\qquad
\text{FDR}=\frac{\text{FP}}{\text{CL+FP}},
\end{equation}
where by convention we take $\text{FDR}=0$ if its numerator and denominator are both zero, i.e., if there are no estimated links.
The recall is defined to be
$$\text{recall} = 1-\text{FNR}.$$
The precision is defined to be
$$\text{precision} = 1-\text{FDR}.\footnote{Note that the precision for a blocking procedure is not expected to be high since we are only placing similar pair in the same block (not not fully running an entity resolution procedure or de-duplication procedure, which would try and maximize both the recall and the precision).}$$ The reduction ratio (RR) is defined as 
$$\text{RR} = 1 - \frac{s_\text{M} + s_\text{N}}{n_\text{M}+n_\text{N}},$$ where $n_M$ and $n_N$ are the total of matched and non-matched records and the number of true matched and true non-matched candidate record pairs generated by an indexing technique is denoted with
$s_\text{M} + s_\text{N} \leq n_\text{M}+n_\text{N}.$ 
The RR provides information about how many candidate record pairs were generated by an indexing technique compared to all possible record pairs, without assessing the quality of these candidate record pairs. We also evaluate the methods using the precision, where  precision calculates the proportion of how many of the classified matches (true positives + false positives) have been correctly classified as true matches (true positives). It thus measures how precise a classifier is in classifying true matches. This measure is useful if we wish to use hashing based approaches for entity resolution, however, as we show, we are not able to achieve both a high precision and recall (see \S \ref{sec:app}). It's most important for a blocking method to have a high RR and recall because the entity resolution task can correct for potential problems that are represented with a low precision. On the other hand, the error summarized by the recall cannot be improved by an entity resolution task. 

\section{Application}
\label{sec:app}

We test the two blocking approaches on a subset of the ongoing Syrian conflict, where via the Human Rights Data Analysis Group (HRDAG), we have access to four databases from the Syrian conflict which cover roughly the same period, namely March 2011 -- April 2014. In this section, we apply LSH based methods to the subset of the Syrian dataset. (We do not consider any methods in the literature that performed worse than KLSH in terms of RR and recall. See \cite{sadosky2015blocking} for further details and experiments on traditional and other probabilistic blocking schemes.)

\subsection{The Syrian Data}
The four data sources consist of the Violation Documentation Centre (VDC), Syrian Center for Statistics and Research (CSR-SY), Syrian Network for Human Rights (SNHR), and Syria Shuhada website (SS). Each database lists a different number of recorded victims killed in the Syrian conflict, along with available identifying information including full Arabic name, date of death, death location, and gender. 
Since the above information is collected indirectly, such as through friends and religious leaders, or traditional media resources, it comes with many challenges. 
For example, the data set contains natural biases, spelling errors, missing values in addition to duplication of those killed in the conflict. The ambiguities in Arabic names make the situation more challenging as there can be a large textual difference between the full and short names in Arabic. Such ambiguities and lack of additional information make blocking on this data set considerably challenging~\citep{price2014updated}. Owing to the significance of the problem, HRDAG has provided labels for a large subset of the data set. More specifically, five different human experts from the HRDAG manually reviewed pairs of records in the four data sets, classifying them as matches if referred to the same entity and non-matches otherwise.  (More information regarding the Syrian data set can be found in Appendix \ref{sec:syrian}).

\subsection{KLSH applied to Syrian data}
We first apply KLSH to the subset of the Syrian data set, which greatly contrasts the empirical studies shown in \cite{steorts14comparison}. The parameters to be set for KLSH are the number of random projections (p) and the number of clusters to output (k). Using this k-means approach to blocking, the mean number of records within a cluster can be fixed.

\begin{figure}[htbp]
\begin{center}
\includegraphics[width=0.45\textwidth]{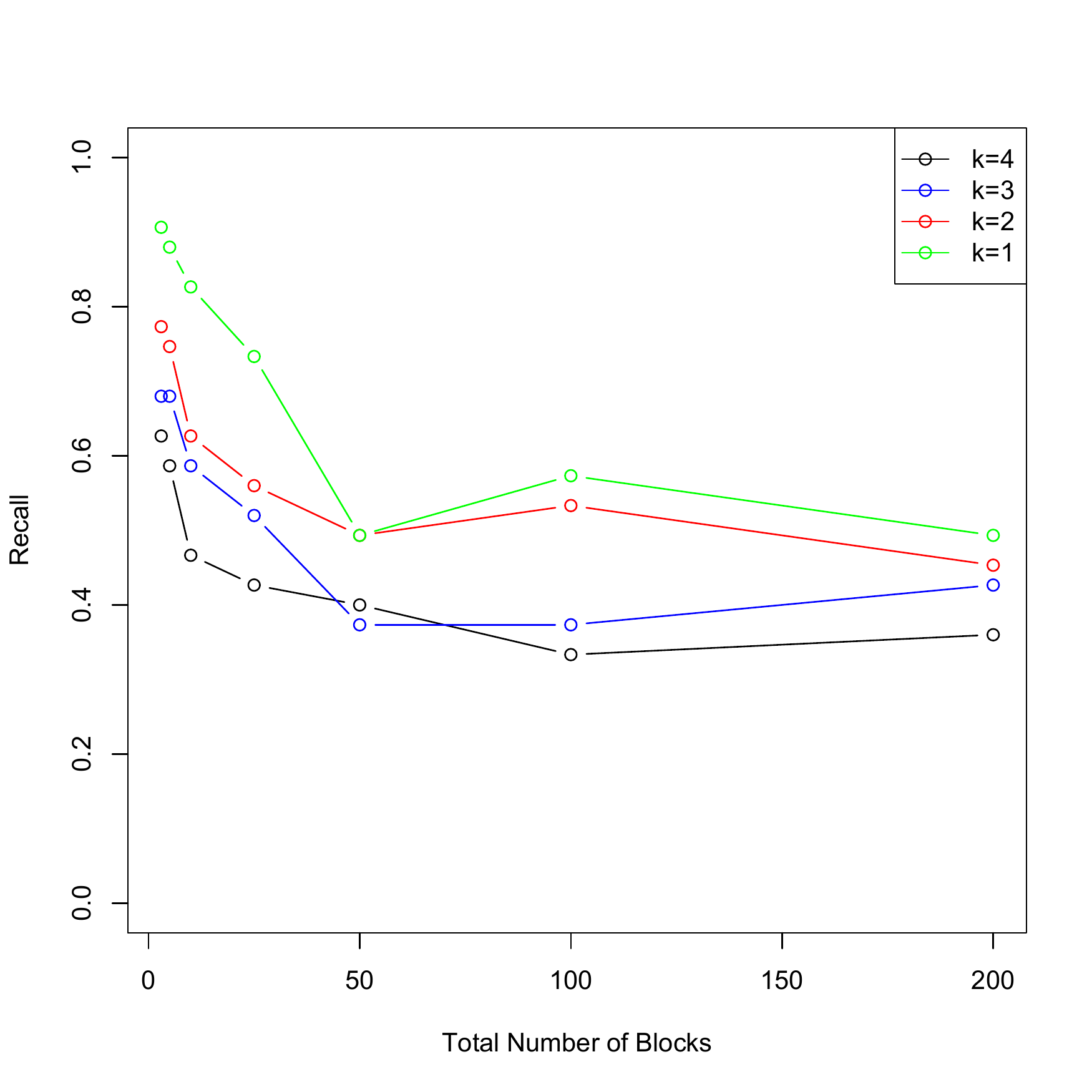}
\includegraphics[width=0.45\textwidth]{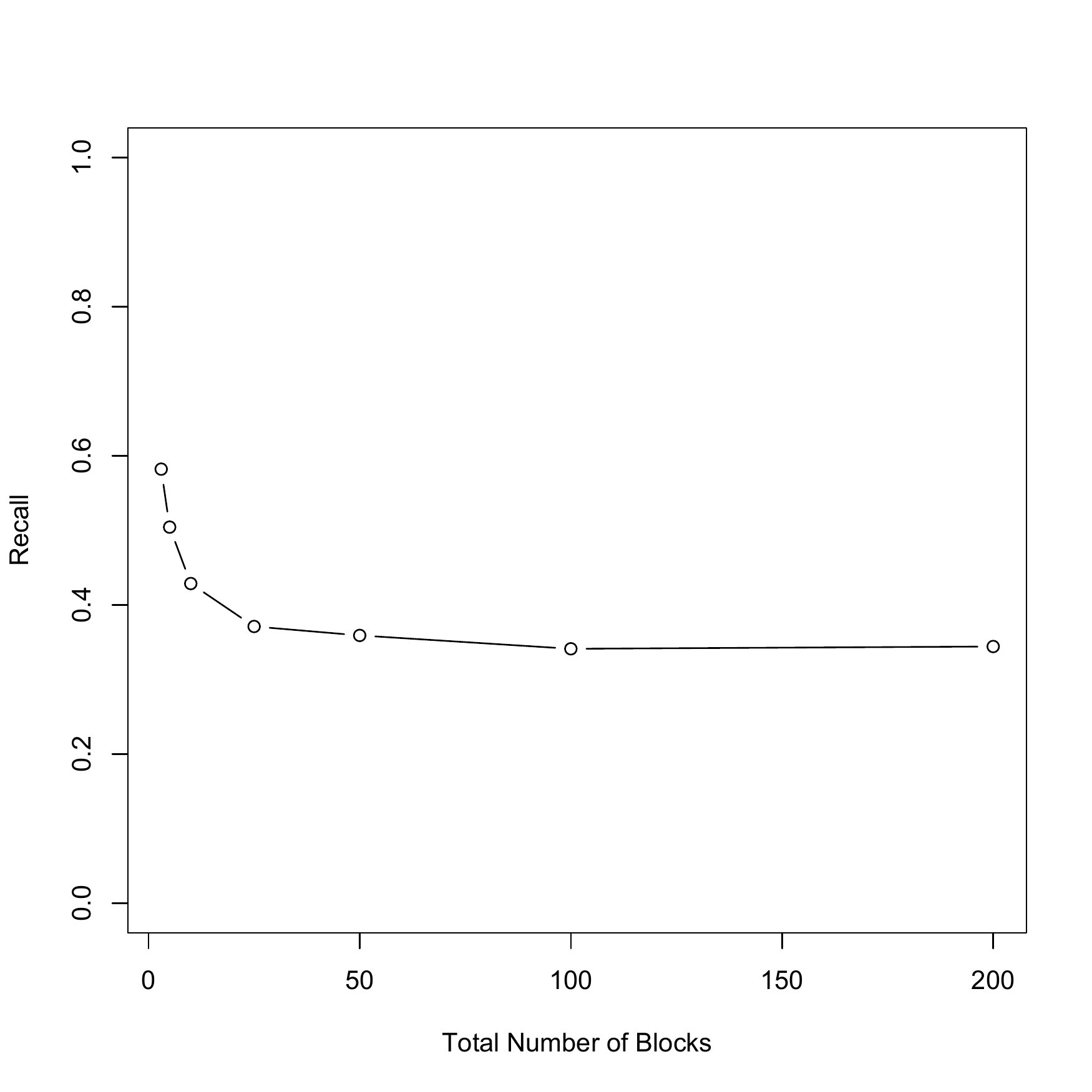}
\caption{Left: KLSH on subset of Syria database (20,000 records) using p=20. Right: KLSH on entire Syrian database using p=20 and k=1. One can see that the recall is very poor compared with previous approaches applied using KLSH, and thus, the method is not suitable for blocking on this particular data set.
}
\label{fig:klsh-subset}
\end{center}
\end{figure}

Figure \ref{fig:klsh-subset} (left panel) displays the results of KLSH clustering applied on the subset of the Syrian database, where we plot the recall versus the total number of blocks. We set the number of random projections to be $p=20$ and allow the shingles to vary from $k=1,2,3,4.$ This figure shows that a 1-shingle always achieves the highest recall. We notice that using a 1-shingle, a block size of 100, the recall is 0.60, meaning that 40\% of the time the same two records are split across blocks. 

\subsection{DOPH applied to Syrian data}
\begin{figure}[htbp]
\begin{center}
\includegraphics[width=0.75\textwidth]{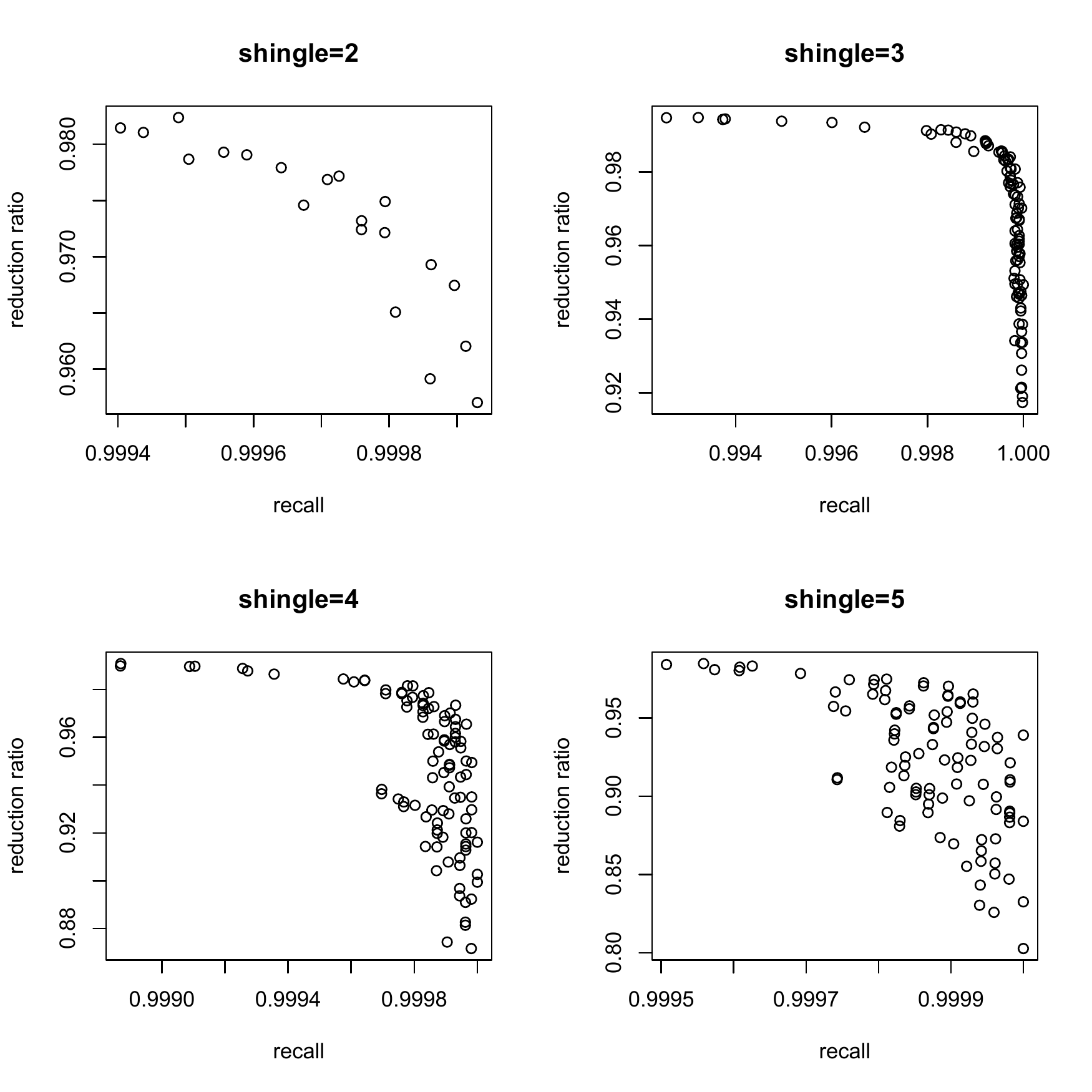}
\caption{For shingles 2--5, we plot the RR versus the recall. Overall, we see the best behavior for a shingle of 3, where the RR and recall can be reached at 0.98 and 1, respectively. We allow L and K to vary on a grid here. L varies from 100--1000 by steps of 100 and K takes values 15,18,20,23,25,28,30,32,35.}
\label{syria-takethatAssad}
\end{center}
\end{figure}

\begin{figure}[htbp]
\begin{center}
\includegraphics[width=0.75\textwidth]{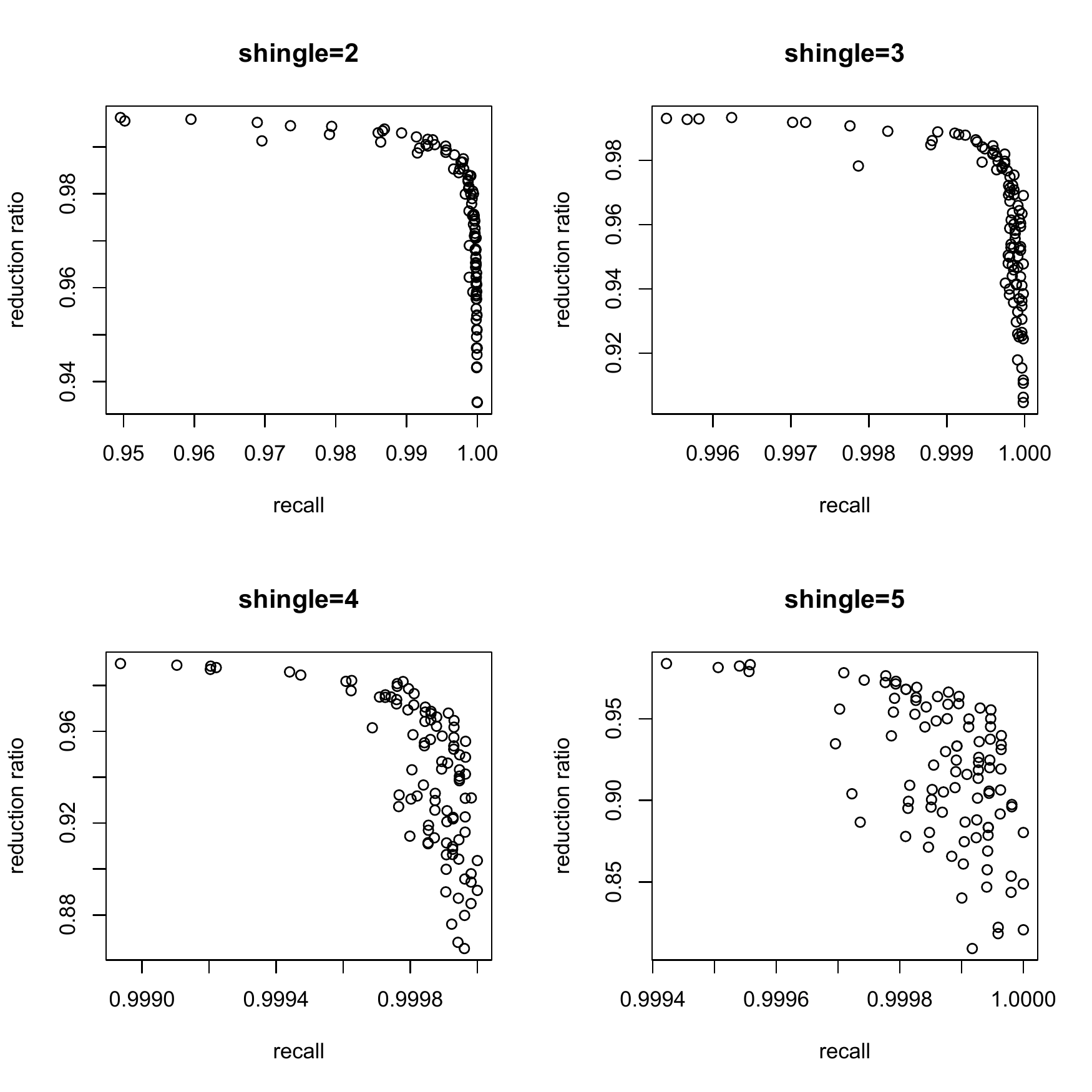}
\caption{For shingles 2--5, we plot the RR versus the recall. Overall, we see the best behavior for a shingle of 2 or 3, where the RR and recall can be reached at 0.98 and 1, respectively. We allow L and K to vary on a grid here. L varies from 100--1000 by steps of 100 and K takes values 15,18,20,23,25,28,30,32,35.}
\label{syria-takethatAssadAgain}
\end{center}
\end{figure}

Due to the poor results achieved by KLSH for the Syrian data set, we apply minhashing using both the unweighted and weighted DOPH algorithm to the full Syrian database using shingles 2---5. We illustrate that regardless of the number of shingles used, the recall and RR are close to 1 as illustrated in Figure~\ref{syria-takethatAssad}. Furthermore, using unweighted DOPH, we see that a shingle of three overall is most stable in having a recall and RR close to 0.99 as illustrated in Figure~\ref{syria-takethatAssadAgain}. Using weighted DOPH, we see that a shingle of two or three overall is most stable in having a recall and RR close to 0.99. In terms of computational run time, we note that each individual run takes 10 minutes on the full Syrian dataset and 100 GB of RAM. We contrast this with the other blocking runs that on 20,000 records from Syria take many hours or 1-2 days (or a week) and return a recall and RR that is unacceptable for entity resolution tasks. 
 
\section{Discussion}
\label{sec:disc} 
We have reviewed two modern approaches for blocking, namely KLSH and DOPH and applied both to a subset of the Syrian conflict. We find that while KLSH has been able to handle data sets with low noise and distortions, it is not able to achieve a high recall on the Syrian data set, and thus, is not a suitable for entity resoultion for data sets that have similar levels of noise as in the Syrian data set. On the other hand, DOPH performs well given the sparsity and noisy levels on the observed data at hand, and appears to be an excellent, stable, and scalable choice for the blocking step in an entity resolution task. This merits further investigations with scalable variants of LSH for entity resolution tasks. 



\section*{Acknowledments} We would like to thank HRDAG for providing the data and for helpful conversations. We would also like to thank Stephen E. Fienberg and Lars Vilhuber for making this collaboration possible.  Steorts's work is supported by NSF-1652431 and NSF-1534412. Shrivastava's work is supported by NSF-1652131 and NSF-1718478.  This work is representative of the author's alone and not of the funding organizations.

\small{
\bibliographystyle{ims}
\bibliography{entity}
}

\clearpage
\newpage
\normalsize
\appendix

\section{Syrian Data Set}
\label{sec:syrian}
In this section, we provide a more detailed description about the Syrian data set. As already mentioned, via collaboration with the Human Rights Data Analysis Group (HRDAG), we have access to four databases. They come from the Violation Documentation Centre (VDC), Syrian Center for Statistics and Research (CSR-SY), Syrian Network for Human Rights (SNHR), and Syria Shuhada website (SS). Each database lists each victim killed in the Syrian conflict, along with identifying information about each person (see \cite{price_2013} for further details).

Data collection by these organizations is carried out in a variety of ways. Three of the groups (VDC, CSR-SY, and SNHR) have trusted networks on the ground in Syria.  These networks collect as much information as possible about the victims. For example, information is collected  through direct community contacts. Sometimes information comes from a victim's friends or family members. Other times, information comes from religious leaders, hospitals, or morgue records.  These networks also verify information collected via social and traditional media sources.  The fourth source, SS, aggregates records from multiple other sources, including NGOs and social and traditional media sources (see \url{http://syrianshuhada.com/} for information about specific sources).

These lists, despite being products of extremely careful, systematic data collection, are not probabilistic samples \citep{Sig2015, IOAS2015, CJLS2015, price2014updated}. Thus, these lists cannot be assumed to represent the underlying population of all victims of conflict violence.  Records collected by each source are subject to biases, stemming from a number of potential causes, including a group's relationship within a community, resource availability, and the current security situation.  


\subsection{Syrian Handmatched Data Set}
\label{app:hand}
We describe how HRDAG's training data on the Syrian data set was created, which we use in our paper. 


First, all documented deaths recorded by any of the documentation groups were concatenated together into a single list. From this list, records were broadly grouped according to governorate and year. In other words, all killings recorded in Homs in 2011 were examined as a group, looking for records with similar names and dates.

Next, several experts review these ``blocks", sometimes organized as pairs for comparison and other times organized as entire spreadsheets for review.  These experts determine whether pairs or groups of records refer to the same individual victim or not. Pairs or groups of records determined to refer to the same individual are assigned to the same ``match group." All of the records contributing to a single ``match group" are then combined into a single record. This new single record is then again examined as a pair or group with other records, in an iterative process.

For example, two records with the same name, date, and location may be identified as referring to the same individual, and combined into a single record. In a second review process, it may be found that that record also matches the name and location, but not date, of a third record. The third record may list a date one week later than the two initial records, but still be determined to refer to the same individual. In this second pass, information from this third record will also be included in the single combined record.

When records are combined, the most precise information available from each of the individual records is kept. If some records contain contradictory information (for example, if records A and B record the victim as age 19 and record C records age 20) the most frequently reported information is used (in this case, age 19). If the same number of records report each piece of contradictory information, a value from the contradictory set is randomly selected.

Three of the experts are native Arabic speakers; they review records with the original Arabic content. Two of the experts review records translated into English. These five experts review overlapping sets of records, meaning that some records are evaluated by two, three, four, or all five of the experts. This makes it possible to check the consistency of the reviewers, to ensure that they are each reaching comparable decisions regarding whether two (or more) records refer to the same individual or not.

After an initial round of clustering, subsets of these combined records were then re-examined to identify previously missed groups of records that refer to the same individual, particularly across years (e.g., records with dates of death 2011/12/31 and 2012/01/01 might refer to the same individual) and governorates (e.g., records with neighboring locations of death might refer to the same individual).

\end{document}